\documentclass[%
 aip,
 amsmath,amssymb,
 reprint,%
]{revtex4-1}

\usepackage{graphicx}
\usepackage{dcolumn}
\usepackage{bm}

\usepackage[utf8]{inputenc}
\usepackage[T1]{fontenc}
\usepackage{mathptmx}
\usepackage{xcolor}

\begin{document}


\title{Amplification of mid-infrared lasers via magnetized-plasma coupling}

\author{Yuan Shi}
\email{shi9@llnl.gov}
\affiliation{Lawrence Livermore National Laboratory, Livermore, California 94550, USA}
 
\author{Nathaniel J. Fisch}%
\affiliation{\mbox{Department of Astrophysical Sciences, Princeton University, Princeton, New Jersey 08544, USA}}%
\affiliation{Princeton Plasma Physics Laboratory, Princeton, New Jersey 08543, USA}

\date{\today}

\begin{abstract}
Plasmas may be used as gain media for amplifying intense lasers, and external magnetic fields may be applied to improve the performance. For mid-infrared lasers, the requisite magnetic field is on megagauss scale, which can already be provided by current technologies. Designing the laser amplifier requires knowing the magnetized three-wave coupling coefficient, which is mapped out systematically in this paper.
By numerically evaluating its formula, we demonstrate how the coupling depends on angle of wave propagation, laser polarization, magnetic field strength, plasma temperature, and plasma density in the backscattering geometry.
Since the mediation is now provided by magnetized plasma waves, the coupling can differ significantly from unmagnetized Raman and Brillouin scatterings.
\end{abstract}

\maketitle

\section{\label{sec:intro}Introduction}
Sources of intense lasers are nowadays available only in the \mbox{$\sim 1\,\mu$m} range via rare-earth-doped solid-state media, and in the \mbox{$\sim 10\,\mu$m} range via carbon-dioxide-based gas mixtures. In the mid-infrared range, windows for atmospheric transmission still exist, but the available laser sources are limited.
To fill in this spectral gap, one has to rely on frequency conversion \cite{Andreev1999co2,Mitrofanov2015mid}, stretching the limits of fiber \cite{Jackson2012towards} and gas \cite{Hassan2016cavity} lasers, or developing photonic crystals as alternative lasing media \cite{Yao2012mid}.
However, these techniques are ultimately limited in intensity by the damaging threshold of optical components, which is typically below the $\sim 100\,\text{MW/cm}^2$ range.
For higher-intensity applications, techniques for further amplifying mid-infrared lasers remain to be developed.

A viable pathway for amplifying mid-infrared lasers beyond the damaging threshold of conventional media is plasma-based laser pulse compression \cite{Malkin1999fast,Andreev2006short}. Instead of relying on cations, molecules, or meta-structures, plasma amplifiers rely on collective modes in already-ionized media to provide resonance.
In unmagentized plasmas, the only two modes are the Langmuir wave, which mediates Raman scattering, and the acoustic wave, which mediates Brillouin scattering. However, once the plasma becomes magnetized, many more waves become available. For example, the upper-hybrid wave \cite{Jia2017kinetic} and kinetic magnetohydrodynamics waves \cite{Edwards2018laser} have been shown to provide effective coupling for \mbox{$\sim 1\,\mu$m} lasers.
The requisite magnetic field is smaller and therefore more readily attainable when amplifying mid-infrared lasers.

In order for the external magnetic field to play a substantial role, the electron-cyclotron frequency should be a non-negligible fraction of the laser frequency. 
For example, the frequency of 5-$\mu$m light is about \mbox{377 Trad/s}. The required magnetic field $B_0$ is thereof \mbox{$\sim 1$ MG}, wherein the cyclotron frequency of an undressed electron is \mbox{$\Omega_e=eB_0/m_e\approx 18$ Trad/s}.
While steady-state megagauss fields may be challenging, they are now readily available via pulsed magnets, which are quasi-static on the time scale of the laser pulse. Using nondestructive magnets, fields on $\sim 1$ MG level are routinely generated at user facilities worldwide \cite{Zherlitsyn2012status,Nguyen2016status,Ding2018construction,Beard2018design}. 
Moreover, with destructive magnets, much higher fields have been obtained using laser-driven targets \cite{Fujioka2013kilotesla,Zhu2015strong,Santos2015laser,Law2016direct,Goyon17} and magnetic flux compression \cite{Gotchev09,Knauer2010compressing,Yoneda12}.

To give a sense of what output intensity might be attainable using plasmas, consider upper-hybrid mediation \cite{Shi2017laser} as an example.
Suppose we use the most intense pump allowed by wavebreaking to amplify the seed pulse.  The highest output intensity is then limited by the modulational instability, where an overcritical intensity can be attained so long as not very many instability growth times are exceeded \cite{Malkin1999fast}. In that case, the unfocused output intensity of mid-infrared pulses can be as large as \mbox{$\sim10^{15}\,\text{W/cm}^2$}. 
In comparison, current technique for amplifying intense mid-infrared pulses relies on difference-frequency
generation using near-infrared optical parametric amplifiers, and subsequent chirped pulse amplification of the idler \cite{Kanai17}. The envisioned output intensity is \mbox{$\sim10^{10}\,\text{W/cm}^2$},
which is limited primarily by the damaging threshold of amplifier crystals \cite{Yin2017towards}. Although the limit imposed by grating damage is somewhat higher \cite{Stuart1995laser}, the solid-state technique is fundamentally limited by material-breakdown intensities, whereas the plasma-based technique is limited by the modulational instability, which only appears at relativistic intensities.

Designing parametric laser amplifier requires knowing the second-order nonlinear susceptibility of the media, which has been studied systematically only for unmagnetized plasmas. When the plasma becomes magnetized, the external magnetic field introduces an optical axis, and the coupling becomes anisotrpic. A practicable formula for the coupling coefficient becomes available only recently, which is obtained by solving the fluid-plasma model to second order in the presence of a background magnetic field \cite{Shi2017three,Shi2019three}. 
Using this formula, the goal of this paper is to present the dependencies of the coupling coefficient on magnetic field strengths and angles of wave propagation under a range of plasma conditions. Mapping out the previously-unknown coupling coefficient will serve as the basis for future designs of plasma-based laser amplifiers.

This paper is organized as follows. In Sec.~\ref{sec:formula}, the analytic formula for the coupling coefficient is briefly reviewed, and its scaling property is articulated. In Sec.~\ref{sec:evaluation}, the formula is evaluated numerically in the backscattering geometry, under conditions relevant to amplifying mid-infrared lasers. Our results are discussed in Sec.~\ref{sec:discussion}, followed by a summary.

\section{Analytic formula \label{sec:formula}}
The coupling coefficient is an essential parameter in the three-wave equations, which can be used to describe wave-wave interactions in nonlinear media. Denoting $a_1, a_2$, and $a_3$ the normalized electric-field envelopes of the pump laser, the signal laser, and the idler wave, respectively. Then, the three-wave equations can be written as $d_t a_1=-\Gamma a_2 a_3/\omega_1$ and $d_t a_{2,3}=\Gamma a_1 a_{3,2}^*/\omega_{2,3}$, where $\omega_j$ is the carrier frequency of $a_j$, and $d_t=\partial_t+\mathbf{v}_g\cdot\nabla+\nu$ is the convective derivative with damping rate $\nu_j$ and wave group velocity $\mathbf{v}_{gj}=\partial\omega_j/\partial\mathbf{k}_j$. In order to resonantly interact, $\omega_1=\omega_2+\omega_3$ and $\mathbf{k}_1=\mathbf{k}_2+\mathbf{k}_3$ satisfy energy-momentum conservation.
In nonrelativistic warm-fluid plasma, the resonant coupling coefficient is \cite{Shi2019three}
\begin{equation}
    \label{eq:coupling}
    \Gamma=\sum_s\frac{Z_s\omega_{ps}^2(\Theta^s+\Phi^s)}{4M_s(u_1u_2u_3)^{1/2}}.
\end{equation}
The underlying ideal warm-fluid model may be well-suited, when wavelengths of interest are much larger than the Debye length $\lambda_D\sim 0.1\,\mu\text{m}\, (T/n)^{1/2}$, while much smaller than the collisional mean free path $\lambda_{\text{mfp}}\sim 100\,\mu\text{m}\, (T^2/n Z^2)$, where the plasma density $n$ is in units of \mbox{$10^{18}\,\text{cm}^{-3}$} and the plasma temperature $T$ is in units of 100 eV.

In the above formula, $Z_s=e_s/e$ and $M_s=m_s/m_e$ are the normalized charge and mass of species $s$, whose plasma frequency is $\omega_{ps}$. In the denominator, $u_j$ is the wave energy coefficient, such that $a_j=eE_ju_j^{1/2}/m_e c\omega_j$ and the averaged wave energy is $\epsilon_0 u_j |E_j|^2/2$. 
In the numerator, $\Theta^s$ is the normalized electromagnetic scattering strength, which equals to the sum of six permutations of $\Theta^s_{1,\bar{2}\bar{3}}$, where $\Theta^s_{i,jl}=(c\mathbf{k}_i\cdot\mathbf{f}_{s,j}) (\mathbf{e}_i\cdot\mathbf{f}_{s,l})/\omega_j$, with $\omega_{\bar{j}}=-\omega_j$, $\mathbf{k}_{\bar{j}}=-\mathbf{k}_j$, and $\mathbf{e}_{\bar{j}}=\mathbf{e}_j^*$. Here, $\mathbf{e}_j$ is the unit polarization vector such that $\mathbf{E}_j=\mathbf{e}_j E_j$, and $\mathbf{f}_{s}=\hat{\mathbb{F}}_{s}\mathbf{e}$, where $\hat{\mathbb{F}}_{s}$ is related to the linear susceptibility by $\chi_{s}=-\omega_{ps}^2 \hat{\mathbb{F}}_{s}/\omega^2$.
Explicitly, the warm-fluid $\hat{\mathbb{F}}_{s}$ is given by the following matrix in the coordinate where $\mathbf{B}_0=(0,0,B_0)$ and $\mathbf{k}=k(\mathrm{s}_\theta,0,\mathrm{c}_\theta)$:
\begin{eqnarray}
    \nonumber
    \hat{\mathbb{F}}\!=\!
    \left(\hspace{-5pt} \begin{array}{ccc}
    \gamma^2(1\!+\!\gamma^2\rho^2\mathrm{s}^2_\theta) \!&\! i\beta\gamma^2(1\!+\!\gamma^2\rho^2\mathrm{s}^2_\theta) \!&\! \gamma^2\rho^2\mathrm{s}_\theta\mathrm{c}_\theta \\
    -i\beta\gamma^2(1\!+\!\gamma^2\rho^2\mathrm{s}^2_\theta)  \!&\! \gamma^2(1\!+\!\beta^2\gamma^2\rho^2\mathrm{s}^2_\theta) \!&\! -i\beta\gamma^2\rho^2\mathrm{s}_\theta\mathrm{c}_\theta \\
    \gamma^2\rho^2\mathrm{s}_\theta\mathrm{c}_\theta \!&\! i\beta\gamma^2\rho^2\mathrm{s}_\theta\mathrm{c}_\theta \!&\! 1\!+\!\rho^2\mathrm{c}^2_\theta
    \end{array}\hspace{-5pt} \right)\!.
\end{eqnarray}
The effect of magnetization enters through $\beta=\Omega/\omega$ and $\gamma^2=1/(1-\beta^2)$, as well as $\rho^2=\hat{\gamma}^2u^2k^2/\omega^2$ with $\hat{\gamma}^2=1/(1-\hat{\beta}^2)$ and $\hat{\beta}^2=\gamma^2u^2k^2(1-\beta^2\mathrm{c}_\theta^2)/\omega^2$. Here, $\Omega_s$ is the gyrofrequency and $u_s$ is the thermal speed of species $s$.
The second term in the numerator of Eq.~(\ref{eq:coupling}) is the normalized thermal scattering $\Phi^s=\Phi^s_0+\Phi^s_{1}+\Phi^s_{\bar{2}}+\Phi^s_{\bar{3}}$, where $\Phi^s_0=-(\xi_s-2)\mu_s^2 (c\mathbf{k}_1\!\cdot\!\mathbf{f}_{s,1})(c\mathbf{k}_2\!\cdot\!\mathbf{f}_{s,2}^*)(c\mathbf{k}_3\!\cdot\!\mathbf{f}_{s,3}^*)/\omega_1\omega_2\omega_3$ and $\Phi^s_j=-\mu_s^2(c\mathbf{k}_j\!\cdot\!\mathbf{f}_{s,1})(c\mathbf{k}_j\!\cdot\!\mathbf{f}_{s,2}^*)(c\mathbf{k}_j\!\cdot\!\mathbf{f}_{s,3}^*)/\omega_1\omega_2\omega_3$. Here, $\xi_s$ is the polytropic index and $\mu_s=u_s/c$. Since $\mu_s\ll 1$ in typical discharges, thermal scattering is usually minuscule.

The coupling coefficient determines the parametric amplification process in both the linear and the nonlinear regimes.
In the linear regime, the pump laser is not yet depleted, and the intensity of the signal laser grows exponentially. In the absence of damping, the growth rate is related to the complex-valued coupling coefficient by $\gamma_0=|\Gamma a_1|/\sqrt{\omega_2\omega_3}$. 
After a few exponentiations, the three-wave interaction enters the pump-depletion regime. In this nonlinear regime, the signal amplitude grows as $a_2\propto \gamma_0 t$, while the signal duration shrinks as $\Delta t_2\propto 1/\gamma_0^2 t$. 
A larger coupling thereof enables more rapid pulse compression within a shorter plasma length.

Instead of relying on plasma density and temperature, the coupling can now be tuned by external magnetic fields to optimize pulse compression.
As an intrinsic measure of the coupling, we can remove the dependence on $a_1$ by comparing $\gamma_0$ with the growth rate of Raman backscattering in an unmagnetized plasma of the same density $\gamma_R=\sqrt{\omega_1\omega_p}|a_1|/2$. The normalized growth rate $\mathcal{M}=\gamma_0/\gamma_R$ is then
\begin{equation}
    \label{eq:growthM}
    \mathcal{M}=2\frac{|\Gamma|}{\omega_p^2}\Big(\frac{\omega_p^3}{\omega_1\omega_2\omega_3}\Big)^{1/2},
\end{equation}
which is proportional to the coupling coefficient up to some kinematic factors. Here, $\omega_p^2=\sum_s\omega_{ps}^2$ is the total plasma frequency. Since $\mathcal{M}$ is more directly related to experimental observable, we will present values of the dimensionless $\mathcal{M}$ instead of values of the coupling coefficient.

In the absence of wave damping, the normalized growth rate is invariant under simultaneous scalings of laser and plasma parameters. When we scale the wave frequency by $\omega\rightarrow\lambda\omega$ and scale the wave vector by $\mathbf{k}\rightarrow\lambda\mathbf{k}$, the linear susceptibility is invariant if we also scale the plasma density by $n_s \rightarrow \lambda^2 n_s$ and scale the magnetic field by $B_0\rightarrow\lambda B_0$ while keeping the plasma temperature and the polytropic index constant. Since the eigenmode structure is invariant, both the wave energy coefficient $u$ and the unit polarization vector $\mathbf{e}$ are unchanged. Moreover, it is easy to see that the electromagnetic scattering $\Theta^s$ and the thermal scattering $\Phi^s$ are also invariant. Consequently, $\mathcal{M}$ calculated for one value of $\omega_1$ is representative for other values of the pump laser frequency, as long as plasma parameters are scaled accordingly.

\begin{figure}[t]
    \centering
    \includegraphics[width=0.42\textwidth]{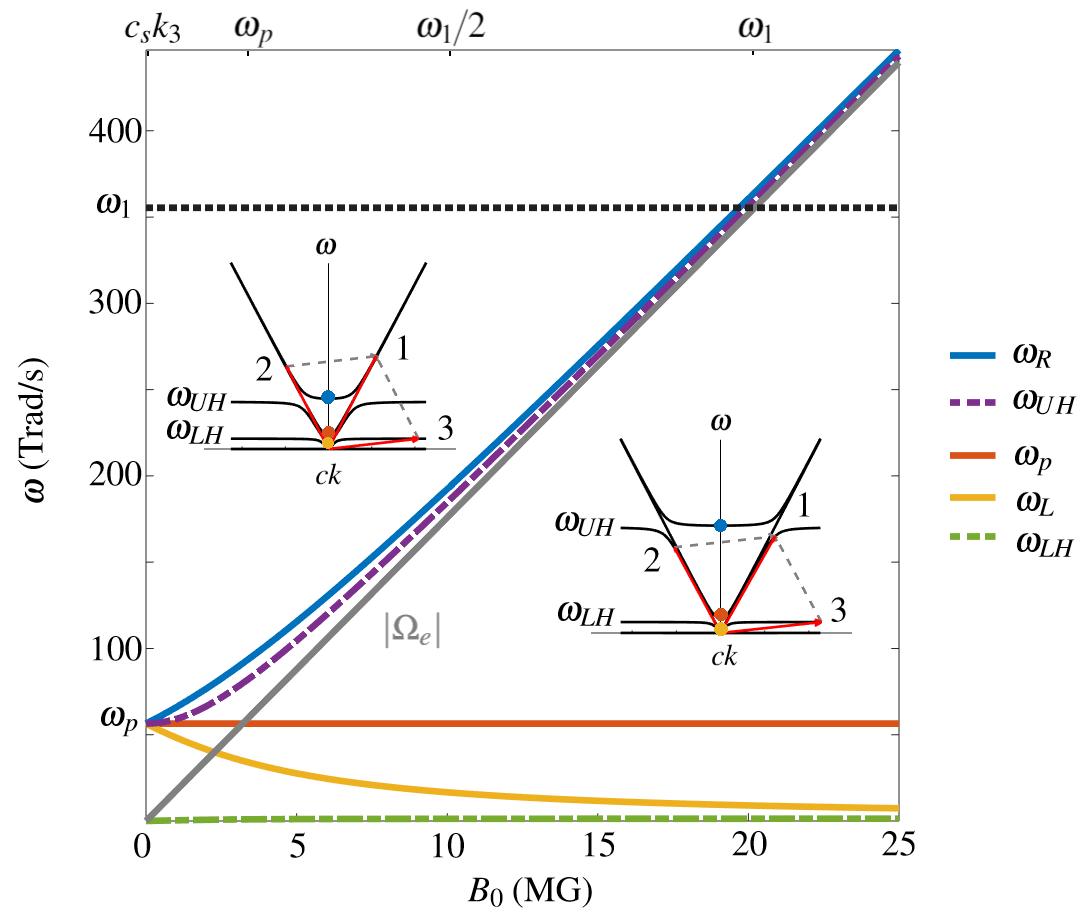}
    \caption{Characteristic frequencies in magnetized hydrogen plasmas with density $n_e=n_i=10^{18}\,\text{cm}^{-3}$ and temperature $T_e=T_i=10$ eV. The cutoff frequencies $\omega_R$ (blue) and $\omega_L$ (yellow) split further from the plasma frequency $\omega_p$ (red) in larger magnetic fields $B_0$. Resonant backscattering with $\omega_1=\omega_2+\omega_3$ and $\mathbf{k}_1=\mathbf{k}_2+\mathbf{k}_3$ can occur both when the pump frequency $\omega_1$ (black) is above cutoff frequencies (left inset) and below resonance frequencies (right inset). In these insets, wave dispersion relations are shown for $\theta_1\lesssim90^\circ$, where the cold-resonance frequencies are $\omega_{UH}$ (purple) and $\omega_{LH}$ (green). The grey line across the diagonal marks the electron-cyclotron frequency. The top axis marks where $|\Omega_e|$ equals to characteristic frequencies of three-wave interactions, where $c_s$ is the sound speed.
    }
    \label{fig:freq}
\end{figure}

As a representative case for mid-infrared lasers, consider 5.3-$\mu$m light, which can be produced either directly by carbon-monoxide lasers or indirectly by frequency doubling of carbon-dioxide lasers. The corresponding pump frequency is \mbox{$\omega_1\approx 355.4$ Trad/s}.
As a comparison, characteristic plasma frequencies as functions of $B_0$ are shown in Fig.~\ref{fig:freq}. The pump laser can propagate either when $\omega_1$ is above cutoff frequencies (left inset), or when $\omega_1$ is below resonance frequencies (right inset). 
In quasi-neutral two-species plasmas, the dispersion relation contains three gapped modes and three gapless modes. 
When $\theta<90^\circ$, the gapped branch with the highest frequency is the right-handed (R) light wave, whose cutoff frequency is $\omega_R$ (solid blue). The second eigenmode is the left-handed (L) light wave, whose cutoff frequency is $\omega_p$ (solid red). The third branch is again left-handed, whose cutoff frequency is $\omega_L$ (solid yellow). The frequency of this branch asymptotes to the upper-hybrid (UH) frequency $\omega_{UH}$ (dashed purple) in the cold limit when $ck\rightarrow\infty$ and $\theta\rightarrow 90^\circ$. 
In the same limit, the gapless mode with the highest frequency symptotes to the lower-hybrid (LH) wave with frequency $\omega_{LU}$ (dashed green), while the frequency of the other two gapless modes approach zero.
The eigenmodes are elliptically polarized except when $\theta=90^\circ$, and the handedness flips for $\theta>90^\circ$. Moreover, magnetized plasma waves are usually neither purely transverse nor purely longitudinal. The waves are more transverse when $\omega$ is closer to $ck$, while more longitudinal when the dispersion relation is further away from the light cone.

For convenience, we will refer to a plasma wave branch by its rough characteristics.
The wave most closely resembles the sound wave is labeled by S. The hybrid wave where ions play and important role is labeled by A. The hybrid wave whose frequency is the closest to electron-cyclotron frequency is labeled by F. Finally, the hybrid wave whose frequency is the closest to the plasma frequency is labeled by P. 
In the magnetohydrodynamics (MHD) limit, S is the slow wave, A is the Alfv\'en wave, F is the fast wave, and hence are these labels. In the opposite limit and when $\theta=0^\circ$, A is the ion-cyclotron wave, F is the electron-cyclotron wave, and P is the Langmuir wave.
On the other hand, when $\theta=90^\circ$, F is the LH (UH) wave and P is the UH (LH) wave when $\omega_p>|\Omega_e|$ ($\omega_p<|\Omega_e|$).
Notice that these labels only give a rough indication of the wave characteristics. The detailed situations may be more complicated, especially when wave branches cross and hybridize.

\section{Numerical evaluation \label{sec:evaluation}}
Using expressions of the linear susceptibility, the analytic formula can be readily evaluated once the resonance conditions are matched. Resonance matching typically requires numerically solving $k_2$ from the equation $\omega_1=\omega_2(k_2\hat{\mathbf{k}}_2)+\omega_3(\mathbf{k}_1-k_2\hat{\mathbf{k}}_2)$. In experiments, the pump frequency $\omega_1$ is given, and $\mathbf{k}_1$ can be solved from the wave dispersion relation for given direction of pump propagation. Also from the dispersion relation, wave frequencies on the right-hand side can be calculated for fixed $\hat{\mathbf{k}}_2$. The equation is then a scalar equation for $k_2>0$, whose roots can be found numerically.

The growth rate depends on the interaction geometry, as well as plasma parameters. The three geometrical degrees of freedom are the angle $\theta_1=\langle \mathbf{k}_1, \mathbf{B}_0\rangle$, the angle $\theta_2=\langle \mathbf{k}_2, \mathbf{B}_0\rangle$, and the angle $\alpha=\langle \mathbf{k}_1, \mathbf{k}_2\rangle$. In this paper, we will focus on the special case $\theta_2=180^\circ-\theta_1$ and $\alpha=180^\circ$, because this backscattering geometry is usually adopted for pulse compression.
In addition, we will further focus on adiabatic hydrogen plasmas in thermal equilibrium with $M_i=1837$, $Z_i=1$, and $\xi_i=\xi_e=3$, for which $n_i=n_e=n_0$ and $T_i=T_e=T_0$. Then, the normalized growth rate is a function of four variables: $\theta_1$, $B_0$, $T_0$, and $n_0$. In the remaining part of this paper, we will give a number of examples to illustrate how $\mathcal{M}$ depends on these four parameters.

\begin{figure}[b]
    \centering
    \includegraphics[width=0.48\textwidth]{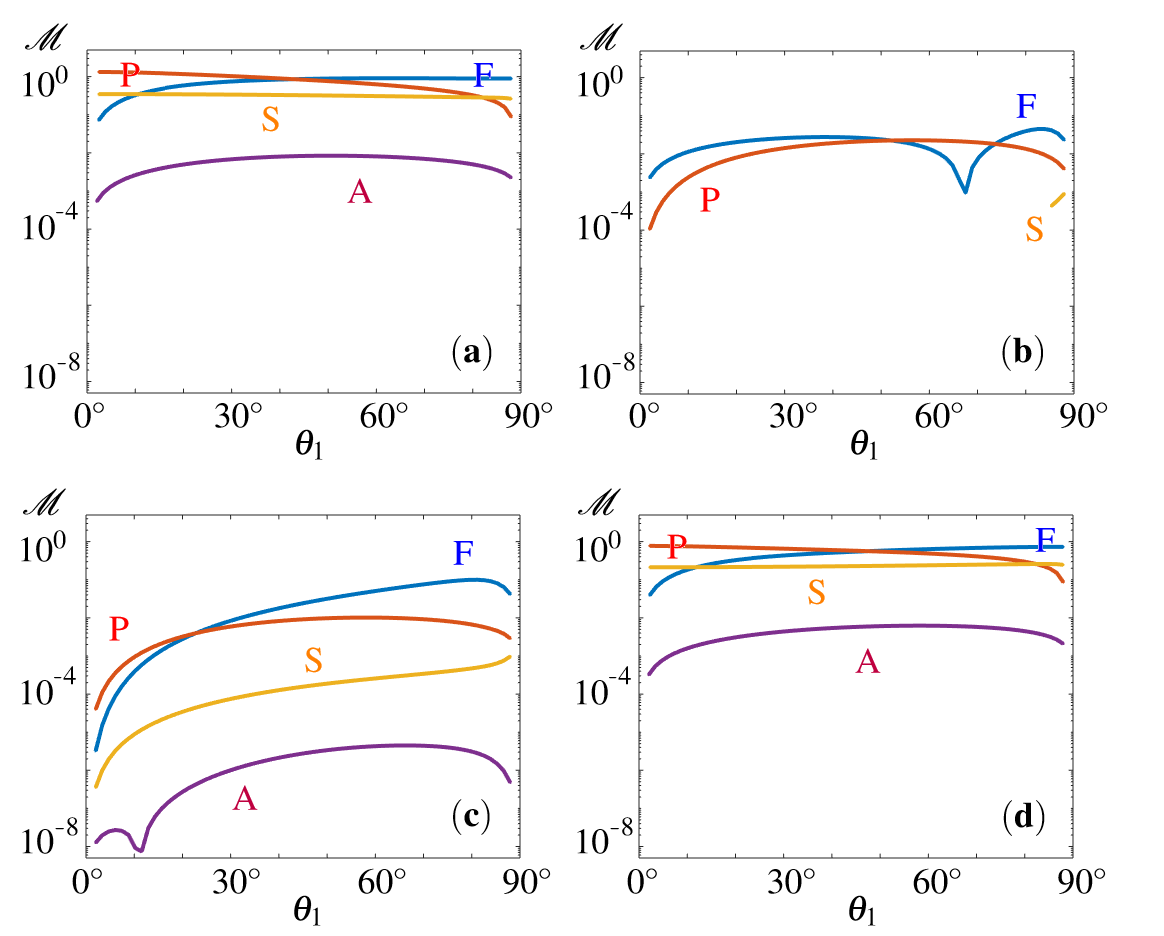}
    \caption{Polarization dependence of $\mathcal{M}$ in a hydrogen plasma with $B_0=5$ MG, $n_0=10^{18}\,\text{cm}^{-3}$, and $T_0=50$ eV. The pump laser, for which \mbox{$\omega_1\approx 355$ Trad/s} and $\theta_1=\langle\mathbf{k}_1,\mathbf{B}_0\rangle$, couples with a backward-propagating signal laser via plasma waves on the P branch (red), the F branch (blue), the S branch (yellow), and the A branch (purple). When $\theta_1<90^\circ$, the R-wave pump couples most strongly with the L wave (a), while the opposite L-R coupling has comparable but smaller growth rates (d). In backscattering geometry, R-R (b) and L-L (c) couplings are polarization-suppressed. Curves are shown only when three-wave resonance conditions can be satisfied.
    }
    \label{fig:pol}
\end{figure}

First, it is important to recognize that $\mathcal{M}$ depends sensitively on the polarization of participating lasers (Fig.~\ref{fig:pol}). Unlike in the unmagnetized case where the two light waves are degenerate, now the two eigenmodes with the same frequency have different $\mathbf{k}$, $\mathbf{e}$, and $\mathbf{f}_s$. 
Specifically for backscattering, couplings between two R waves (Fig.~\ref{fig:pol}b) and two L waves (Fig.~\ref{fig:pol}c) are suppressed, because it is difficult for the plasma wave to carry large angular momentum, which is required to satisfy angular-momentum conservation intrinsic to Eq.~(\ref{eq:coupling}).  On the other hand, cross couplings between R and L waves have little angular momentum discrepancy, so that plasma waves, which are mostly longitudinal, can readily mediate the interaction.
When $\theta_1<90^\circ$, the R-wave pump and L-wave signal (Fig.~\ref{fig:pol}a) have smaller wave vectors and is more strongly coupled to the plasma than the opposite case where the pump is L-wave and the signal is R-wave (Fig.~\ref{fig:pol}d). The growth rates for these later two cases are otherwise qualitatively similar. In what follows, we will focus on R-L scattering, which usually has the largest growth rate.

\begin{figure}[b]
    \centering
    \includegraphics[width=0.45\textwidth]{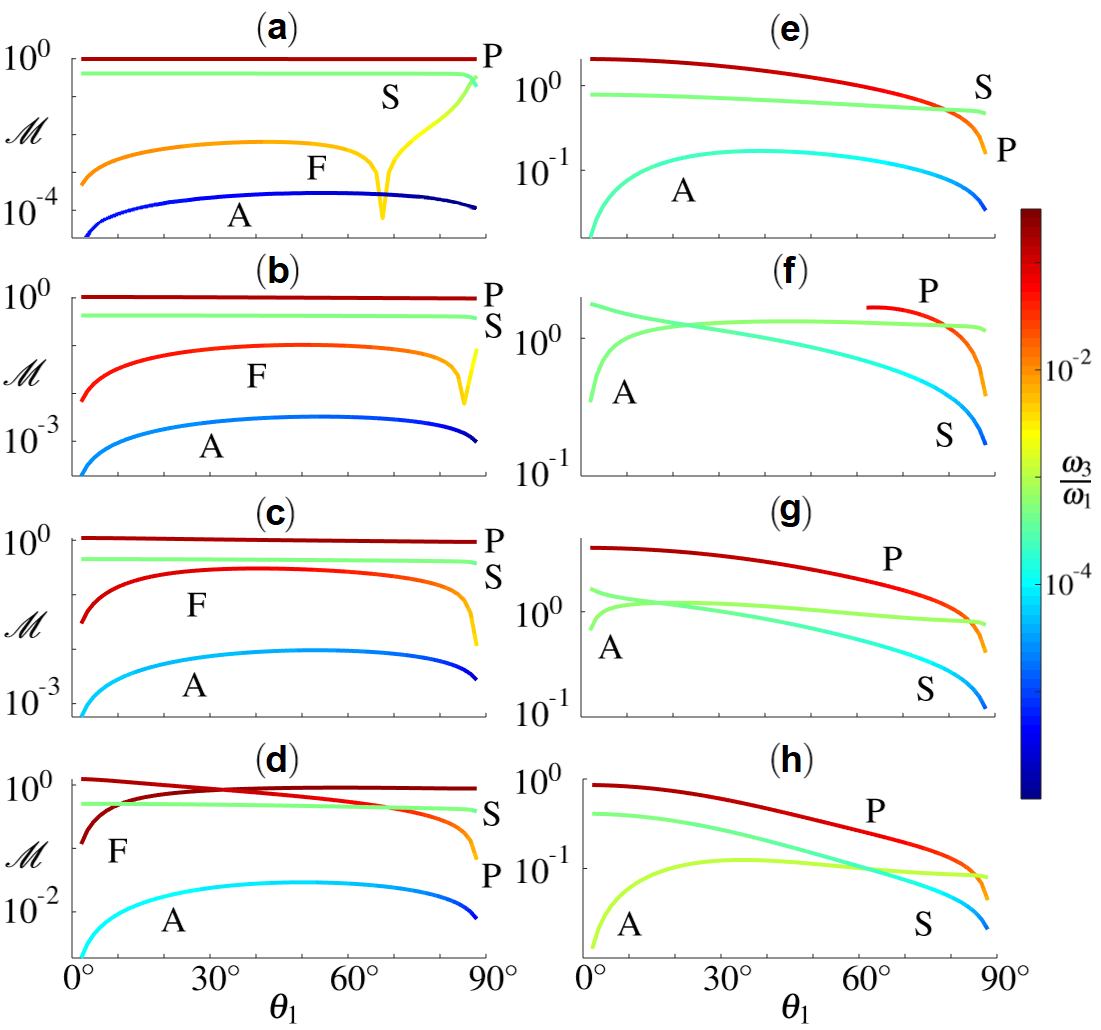}
    \caption{Angle-dependent $\mathcal{M}$ transitions through different regimes as $B_0$ increases, as shown here by a set of examples for backward R-L coupling (\mbox{$\omega_1\approx 355$ Trad/s}) in hydrogen plasmas with $n_0=10^{18}\,\text{cm}^{-3}$ and $T_0=10$ eV. The curves are color-coded by frequencies of the mediating plasma waves. The weakly magnetized case with $c_sk_3>\omega_{LH}$ is exemplified by (a) $B_0=0.2$ MG, where S-F hybridization occurs at large $\theta_1$. When $B_0=1$ MG (b), S-F crossover no longer happens, but F mediation is still polarization suppressed at special angles, which disappears in (c) where $B_0=2$ MG. Magnetization plays important role when $|\Omega_e|>\omega_p$, where the P and F branches cross, as shown in (d) where $B_0=4$ MG. Once $|\Omega_e|>\omega_1/2$, as is the case when $B_0=10$ MG (e), the F resonance is lost. When $B_0=18$ MG (f), $\omega_R$ approaches $\omega_1$, and P mediation is allowed only for large $\theta_1$. Incidentally, S-A crossover also occurs. With slightly larger $B_0=22$ MG (g), $\omega_1$ is now below $\omega_{UH}$, and the pump wave switch from the R to the F branch. With further increase of $B_0$, for example, $B_0=40$ MG (h), $\mathcal{M}$ decreases rapidly with little qualitative change until $B_0$ becomes prohibitively large. 
    }
    \label{fig:B0}
\end{figure}

Second, as the magnetic field increases, the growth rate transitions through a number of qualitatively different regimes (Fig.~\ref{fig:B0}). In weak magnetic fields, the P branch is essentially the unmagnetized Langmuir wave, so $\mathcal{M}\sim1$ is close to Raman scattering. Moreover, when the sound frequency $c_sk_3>\omega_{LH}$ (Fig.~\ref{fig:B0}a), the S branch is close to the unmagnetized sound wave for small $\theta_1$, where the scattering is close to Brillouin. However, for larger $\theta_1$, the frequency of the S branch, which is roughly proportional to $\cos\theta_1$, decreases below $\omega_{LH}$, where S and F branches hybridize. Before this happens, the F branch is dominated by electron-cyclotron motion, and the A branch is dominated by ion-cyclotron motion, whose mediations give rise to very small growth rates. The S-F hybridization can no longer occur in larger $B_0$ (Fig.~\ref{fig:B0}b), where the F branch always has higher frequency than the S branch. For given $k_3<2k_1$, the F branch has substantial transverse components in weak fields, so its mediation may be polarization-suppressed at special $\theta_1$. This phenomena no longer occurs for larger $B_0$ (Fig.~\ref{fig:B0}c), beyond which the F branch becomes mostly longitudinal.
For even larger $B_0$ where $|\Omega_e|>\omega_p$ (Fig.~\ref{fig:B0}d), the F branch hybridizes with the P branch. After P-F crossover, the P branch becomes the LH wave while the F branch becomes the UH wave at large $\theta_1$, so $\mathcal{M}$ becomes significantly modified. When $|\Omega_e|$ increases beyond the two-magnon resonance at $\omega_1/2$, the F-wave mediation is lost (Fig.~\ref{fig:B0}e), because resonance conditions can no longer be satisfied.
In even larger magnetic fields, $\omega_R$ approaches $\omega_1$ and S-A hybridization starts to occur. Incidentally, both phenomena are captured in Fig.~\ref{fig:B0}f for this particular example, where \mbox{$n_0=10^{18}\,\text{cm}^{-3}$} and \mbox{$T_0=10$ eV}. Since the wave dispersion relation is strongly modified, resonance conditions cannot be satisfied for small $\theta_1$ via the P-wave mediation. However, once $|\Omega_e|$ increases beyond $\omega_R$, resonance conditions can again be satisfied for P-wave mediation at all $\theta_1$ (Fig.~\ref{fig:B0}g). In this regime, the pump laser, which used to be on R branch (Fig.~\ref{fig:freq}, left inset), is now on the F branch (Fig.~\ref{fig:freq}, right inset).
With further increase of $B_0$, no qualitative change exist until $\Omega_i$ becomes comparable to $\omega_p$, which requires prohibitively large magnetic field for typical gas-jet plasmas. With only quantitative change (Fig.~\ref{fig:B0}h), $\mathcal{M}$ drops rapidly after the pump wave switches to the F branch. Notice that $\mathcal{M}> 1$ when $|\Omega_e|\sim\omega_1$ (Figs.~\ref{fig:B0}f and \ref{fig:B0}g). The growth rates are otherwise smaller or comparable to Raman.

\begin{figure}[t]
    \centering
    \includegraphics[width=0.3\textwidth]{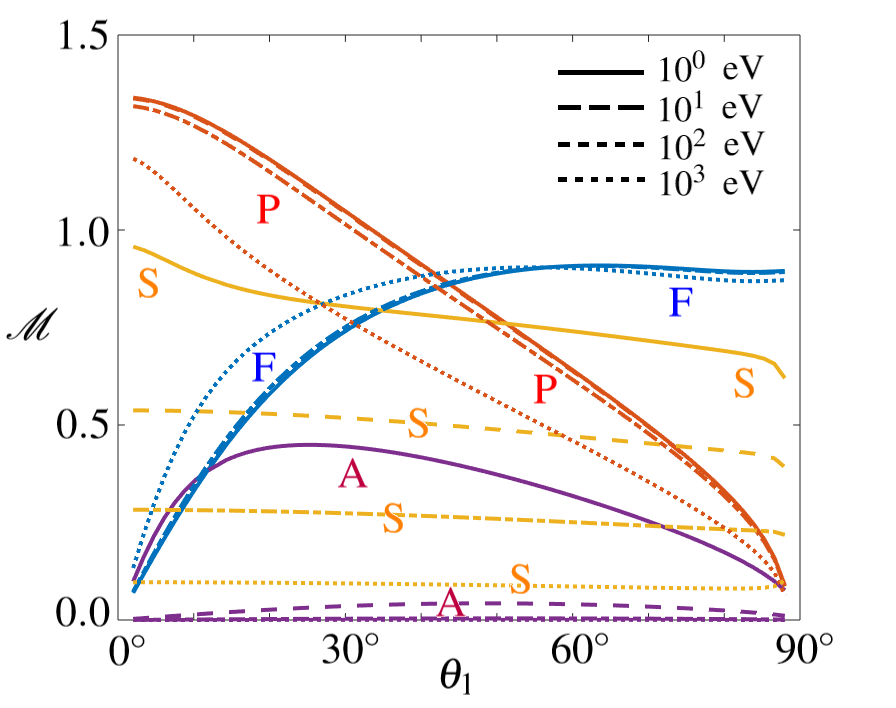}
    \caption{Temperature dependence of backward R-L coupling (\mbox{$\omega_1\approx 355$ Trad/s}) in hydrogen plasmas with \mbox{$n_0=10^{18}\,\text{cm}^{-3}$} at \mbox{$B_0=5$ MG}. Coupling via the P branch (red) and F branch (blue) have weak $T_0$ dependence. On the other hand, coupling via the S branch (yellow) and A branch (purple) are strongly enhanced at low temperature due to S-A hybridization, which vanishes at higher plasma temperature.
    }
    \label{fig:T0}
\end{figure}

Third, the growth rate depends on plasma temperature, as shown in Fig.~\ref{fig:T0} for fixed \mbox{$B_0=5$ MG} and \mbox{$n_0=10^{18}\,\text{cm}^{-3}$}, which is closely related to Fig.~\ref{fig:pol}a and Fig.~\ref{fig:B0}d. Notice that $\mathcal{M}$ is the growth rate excluding wave damping, which has separate dependence on $T_0$.
Since the resonant sound frequency is much less than both $\omega_p$ and $|\Omega_e|$, scattering from the P branch (red) and F branch (blue) have weak dependence on $T_0$. On the other hand, the resonant sound frequency is comparable to or larger than \mbox{$\Omega_i\approx 0.05$ Trad/s}, so scattering from the A branch (purple) and S branch (yellow) have strong temperature dependencies. 
In cold plasma with $T_0=1$ eV (solid), the sound frequency \mbox{$c_sk_3\approx0.06$ Trad/s} is close to $\Omega_i$. Hence, the A and S branches are strongly hybridized. In other words, the A branch also has sound character, while the S branch also has Alfv\'en character. Since both branches have substantial longitudinal and transverse components, they overlap favorably with the transverse lasers, and the coupling is greatly enhanced.
In warmer plasma with $T_0=10$ eV (large dashed), the degree of hybridization is reduced, and the couplings start to return to their bare values.
At even higher temperature \mbox{$T_0=100$ eV} (small dashed) and \mbox{$T_0=1$ keV} (dotted), the two branched become decoupled. The S branch then becomes the unmagnetized sound wave, which gives rise to Brillouin scattering with $\mathcal{M}\sim M_i^{-1/4}$. 
At the same time, the A branches becomes the ion-cyclotron wave, whose mediation has $\mathcal{M}\sim 0$ because the coupling is both polarization suppressed and energy suppressed \cite{Shi2018laser}. The polarization suppression is due to the difficulty for satisfying angular momentum conservation, and the energy suppression is because most wave energy is contained in cyclotron motion.

\begin{figure}[t]
    \centering
    \includegraphics[width=0.48\textwidth]{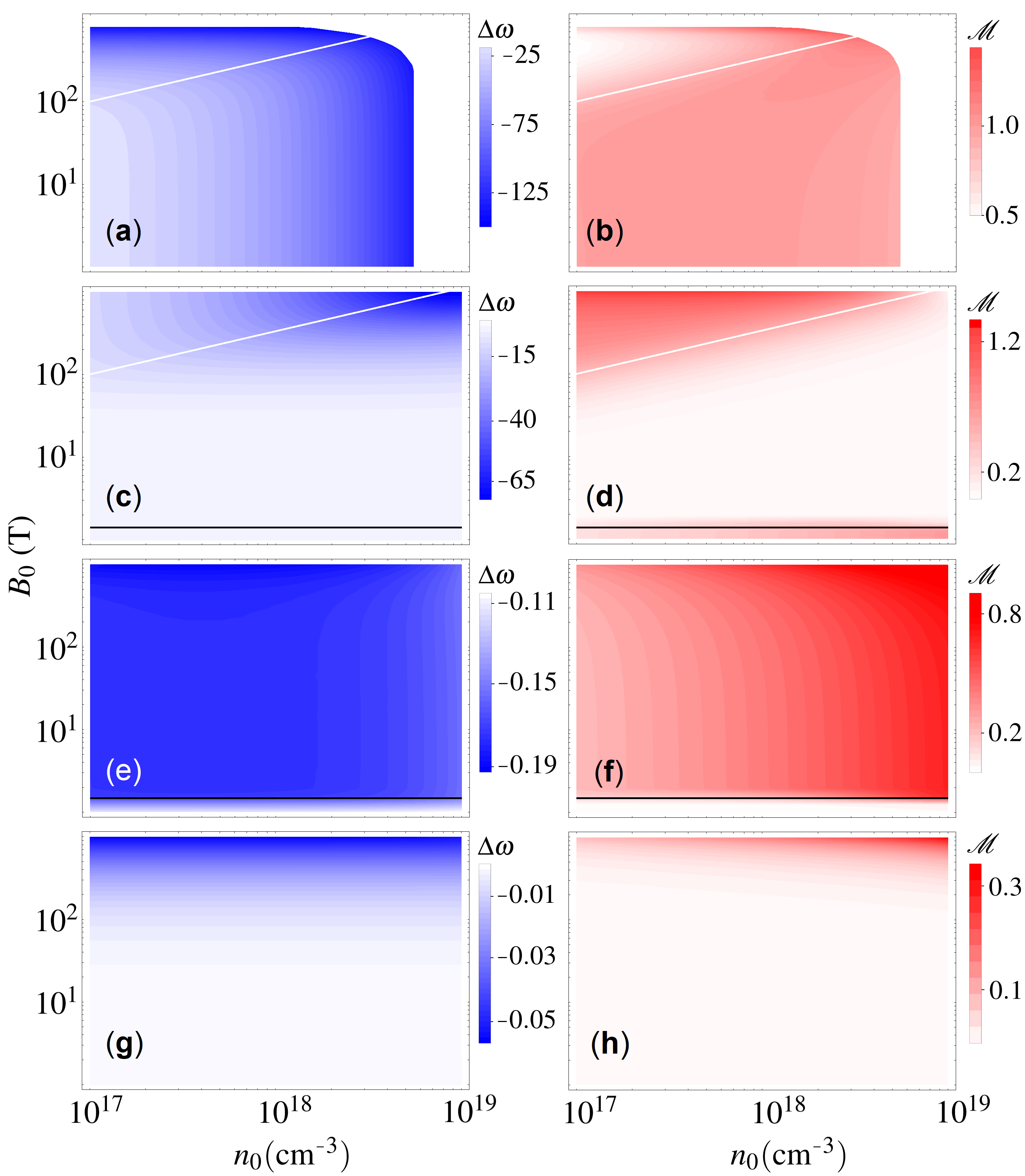}
    \caption{Scan in $n_0$-$B_0$ parameter space with fixed \mbox{$T_0=10$ eV} in a hydrogen plasma for backward R-L coupling when \mbox{$\omega_1\approx 355$ Trad/s} and $\theta_1=50^\circ$. The frequency $\Delta\omega=\omega_2-\omega_1$ is in Trad/s (left), and $\mathcal{M}$ (right) depends most sensitively on the nature of the mediating wave. 
    For the highest-frequency branch (a,~b), the mediation is dominantly via the F (P) wave above (below) the white line. The opposite happens for the second branch (c,~d), which transitions to S-wave mediation below the black line. The S-F transition is flipped for the third branch (e,~f). Finally, for the parameter space shown here, the lowest-frequency branch (g,~h) is always dominated by the A wave.
    }
    \label{fig:nB}
\end{figure}

Finally, the growth rate depends on plasma density. More precisely speaking, $\mathcal{M}$ depends on three dimensionless ratios $\omega_p/\Omega_e$, $\omega_p/c_sk_3$, and $\omega_p/\omega_1$. An example is shown in the $n_0$-$B_0$ parameter space in Fig.~\ref{fig:nB} for fixed $\theta_1=50^\circ$ and \mbox{$T_0=10$ eV}. The frequency downshift $\Delta\omega=\omega_2-\omega_1$ is shown on the left panel, and the normalized growth rate $\mathcal{M}$ is shown on the right panel. 
The highest-frequency branch (Fig.~\ref{fig:nB}a and \ref{fig:nB}b) can no longer mediate laser coupling for large $n_0$ and $B_0$ where $\omega_{UH}\gtrsim\omega_1/2$. Below this threshold, $\Delta\omega$ is close to $\omega_{UH}$, and $\mathcal{M}$ transitions from Raman-type P-wave mediation, which dominates when $|\Omega_e|<\omega_p$ (below the white line), to electron-cyclotron-type F-wave mediation, which dominates when $|\Omega_e|>\omega_p$ (above the white line).
In addition to the P-F transition, the second branch (Fig.~\ref{fig:nB}c and \ref{fig:nB}d) also encounters S-F transition: below the black line, the mediation is via the S wave, while the mediation is via the F wave above the black line. 
The third branch (Fig.~\ref{fig:nB}e and \ref{fig:nB}f) sees the opposite of the S-F transition. With larger $B_0$, the S-wave mediation is further enhanced compared to Brillouin. 
In the parameter space shown here, which is attainable in contemporary experiments, the lowest-frequency branch (Fig.~\ref{fig:nB}g and \ref{fig:nB}h) is always dominated by the A wave. The coupling is weak and increases with $B_0$ as $\Omega_i$ approaches the sound frequency.

\section{Discussion \label{sec:discussion}}
Unlike Raman and Brillouin scattering, magnetized laser coupling has intricate dependencies on laser polarization, interaction geometry, and plasma parameters. Even in the simplest case considered in this paper, namely, the lasers have the polarization of linear eigenmodes, the resonant interaction is in backscattering geometry, and the plasma is constituted of two species in thermal equilibrium, the coupling still depends on four independent variables: $\theta_1$, $B_0$, $T_0$, and $n_0$. Mapping out a complete atlas of this four-dimensional space is already demanding, and additional effects such as damping and nonuniformity will likely further complicate the situation.

Instead of attempting to exhaust all possible scenarios, this paper aims to elucidate the big picture in the weak-coupling regime when the fluid model is applicable. Using a set of examples, the big picture may be summarized by the following rules of thumb. (1) Magnetization starts to affect Brillouin when $\Omega_e\sim c_sk_3$ and starts to affect Raman when $\Omega_e\sim \omega_p$. The coupling is greatly enhanced when $\Omega_e\sim \omega_1$ and subsequently suppressed when $B_0$ becomes much larger. (2) The coupling is usually stronger when the participating waves have mixed characteristics. When the degree of hybridization reduces, the mediating mode with stronger cyclotron character provides weaker coupling. (3) Electron-dominated modes provide larger coupling than ion-dominated modes. The coupling is therefore stronger when laser polarization is better aligned with electron-cyclotron motion.

In addition to mapping out the growth rate, designing the plasma-based laser amplifier also requires knowing how limiting effects, such as wave damping and modulational instability, depend on the oblique magnetic field. While detailed analysis is beyond the scope of this paper, we can make some simplistic estimations for mid-infrared lasers. Assuming upper-hybrid mediation \cite{Shi2017laser}, then in the regime $\Omega_e/\omega_1\sim0.1$, the maximum pump intensity allowed by wavebreaking is \mbox{$\sim10^{13}\,\text{W/cm}^2$}. The maximum amplification time limited by modulational instability is \mbox{$\sim 10^2$ ps}, corresponding to a requisite mediating plasma of \mbox{$\sim1$} cm in length. After nonlinear pulse compression, the output pulse reaches an unfocused intensity of \mbox{$\sim10^{15}\,\text{W/cm}^2$} and a pulse duration of \mbox{$\sim 1$ ps}. While these order-of-magnitude estimations are based on upper-hybrid mediation, they give a rough sense of what magnetized pulse compression might be able to achieve in the mid-infrared range.

In summary, we evaluate the analytic formula for the backscattering growth rate in magnetized plasmas by matching the three-wave resonance conditions numerically. Once the magnetic field becomes non-negligible, the growth rate differs significantly from Raman and Brillouin. In addition to plasma paratemeters, the growth rate now depends sensitively on laser polarization and angle of wave propagation. After scaling plasma parameters accordingly, examples demonstrated in this paper are also representative for other laser wavelengths. The numerical examples and the resultant rules of thumb will facilitate future designs of plasma-based laser pulse compressors.

\begin{acknowledgments}
The authors thank E. Michael Campbell for stimulating discussions. The work was supported in part by DOE NNSA DE-NA0003871, and was performed under the auspices of the U.S. Department of Energy by Lawrence Livermore National Laboratory under Contract DE-AC52-07NA27344 and was supported by the Lawrence Fellowship through LLNL-LDRD Program under Project No. 19-ERD-038. 
\end{acknowledgments}


\providecommand{\noopsort}[1]{}\providecommand{\singleletter}[1]{#1}%

\end{document}